\documentclass[
 reprint,
 amsmath,amssymb,
 aps,
]{revtex4-2}

\usepackage{graphicx}
\usepackage{dcolumn}
\usepackage{bm}
\usepackage{siunitx}

\newcommand{\MHz}{\mega\hertz}
\newcommand{\nm}{\nano\meter}

\newcommand{\ms}{\milli\second}
\newcommand{\mm}{\milli\meter}
\newcommand{\uK}{\micro\kelvin}
\newcommand{\um}{\micro\meter}

\usepackage{comment}
\usepackage[normalem]{ulem} 
\usepackage{mdframed} 
\mdfdefinestyle{highlightstyle}{
    backgroundcolor=red,
    linewidth=0pt, 
    innerleftmargin=5pt,
    innerrightmargin=5pt,
    innertopmargin=5pt,
    innerbottommargin=5pt,
}

\begin{document}

\preprint{APS/123-QED}

 \title{Quantifying Light-assisted Collisions in Optical \\  Tweezers Across the Hyperfine Spectrum } 

\author{Steven K. Pampel$^*$, Matteo Marinelli, Mark O. Brown, Jos\'e P. D'Incao, Cindy A. Regal$^{\dagger}$}

\affiliation{%
 $^\mathrm{1}$JILA, National Institute of Standards and Technology and University of Colorado,
and Department of Physics, University of Colorado, Boulder, Colorado 80309, USA}
\date{January 10, 2025}

\begin{abstract}
We investigate the role of hyperfine structure in resonant-dipole interactions between two atoms co-trapped in an optical tweezer. Two-body loss rates from light-assisted collisions (LACs) are measured across the $^{87}$Rb hyperfine spectrum and connected to properties of molecular photoassociation potentials via a semi-classical model. To obtain our results, we introduce an imaging technique that leverages repulsive LACs to detect two atoms in a trap, thereby circumventing parity constraints in tweezers. Our findings offer key insights for exploiting hyperfine structure in laser-induced collisions to control cold atoms and molecules in a broad range of quantum science applications.

\end{abstract}
 
\maketitle
Optical tweezers have emerged as a leading platform for controllable many-atom systems, enabling state-of-the-art applications in quantum information and metrology \cite{weiss2017quantum, henriet2020quantum, browaeys2020many, kaufman2021quantum}. An underlying aspect of control within these systems is the process of light-assisted collisions (LACs), wherein two colliding atoms absorb a photon to form a quasi-molecular state \cite{jones2006ultracold}. At large detunings (1\SI{-100}{GHz}), LACs have been used for bound-state molecular photoassociation \cite{wang2004photoassociative, drag2000experimental, weyland2021pair, picard2023high} and direct laser cooling to a Bose-Einstein condensate \cite{urvoy2019direct}. At smaller detunings, LACs are routinely harnessed for sub-Poissonian loading of tweezers \cite{grunzweig2010near, schlosser2001sub, sortais2012sub, depue1999unity} and parity imaging in quantum-gas microscopes \cite{bakr2009quantum}. Despite their central role in the latter regime, where hyperfine structure and spontaneous emission introduce new complexities \cite{weiner2003cold}, accurate predictions of quantitative LAC behavior remain elusive. 

\begin{figure}[b]
\includegraphics[width=\columnwidth]
{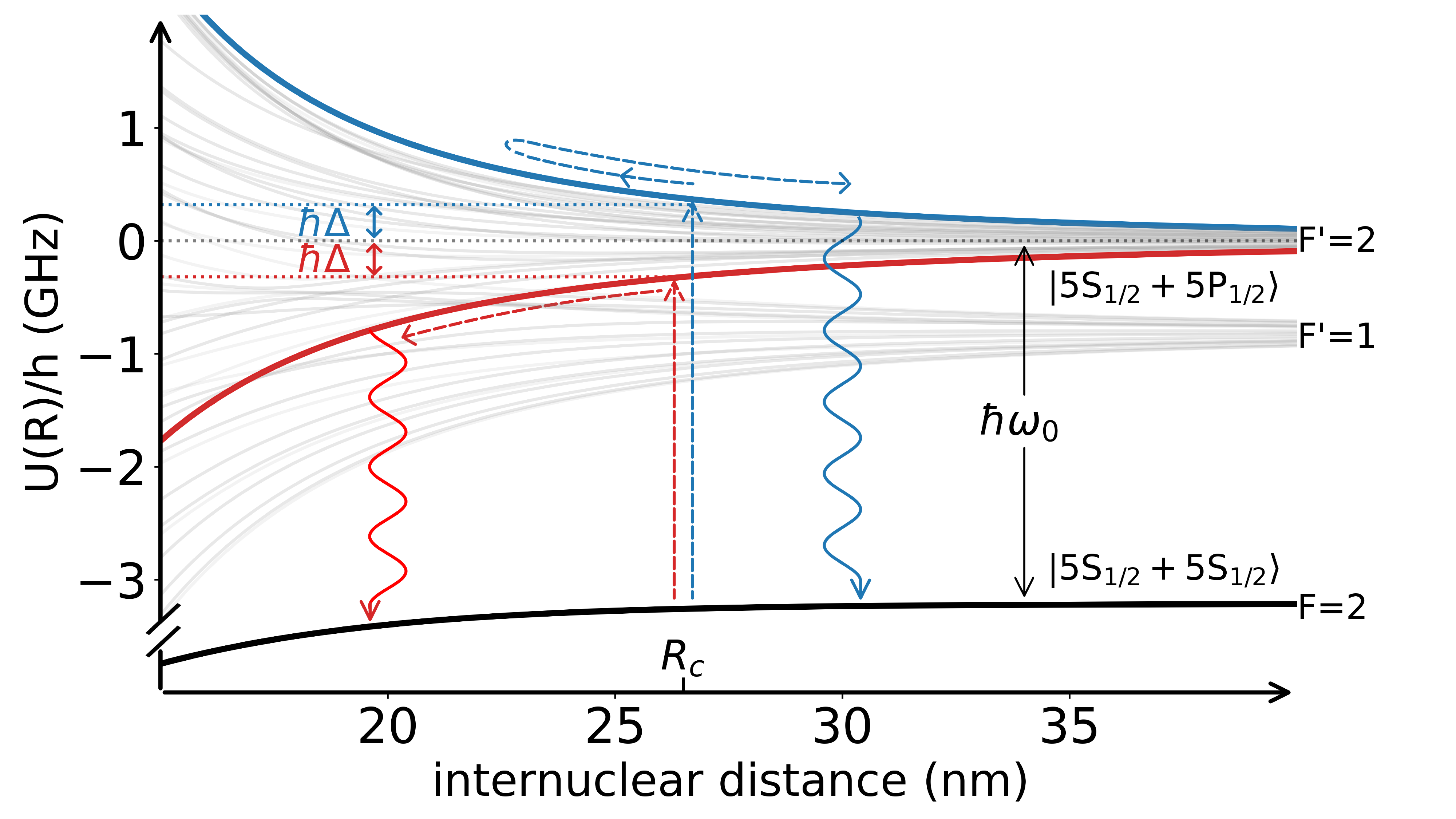}
\caption{\label{fig:photoassociation} Light-assisted collision process on the D1 line of $^{87}\mathrm{Rb}$. Atoms are excited from $S + S $ to $S +P $ at an internuclear distance $R_C$, which is resonant with a photon detuned by $\Delta$ from the free-space transition frequency $\omega_0$. Hyperfine-resolved molecular potentials (grey lines) are shown with example repulsive (blue) and attractive (red) potentials. Atoms traversing the potential (dashed red or blue arrows) can gain kinetic energy before spontaneously decaying to $S+S$.}
\end{figure}

The LAC process is defined by the excitation of atom pairs from an $S + S $ to $S +P$ electronic state at a resonant internuclear distance $R_C$ (Condon radius). For homonuclear systems, the $S + P$ resonant dipole interaction yields $1/R^3$ attractive or repulsive potentials where the amplitude and sign is determined by molecular symmetries \cite{jones2006ultracold}. Effects from fine and hyperfine interactions, which become important at large and small detunings respectively, introduce additional symmetries that create a multitude of interaction potentials. Atom pairs experiencing an $S+P$ potential can gain kinetic energy during an inelastic LAC before spontaneously decaying back to the $S+S$ ground state (Fig.~\ref{fig:photoassociation}), resulting in loss if the final kinetic energy of the atom(s) exceeds the trap depth.

Two-body losses from LACs were initially studied by measuring the fluorescence decay of an atomic ensemble in a magneto-optical trap (MOT) ~\cite{weiner1999experiments,lett1995hyperfine,wallace1992isotopic,ueberholz2002cold}. While this method  provides valuable insight, the large atom number and limited variability of the MOT beams prevents the study of isolated two-body dynamics over a wide range of collisional light properties. More recent experiments using optical tweezers avoid these constraints, but have focused primarily on single-atom collisional losses that can enable loading probabilities as high as $96$\% \cite{fuhrmanek2012light,grunzweig2010near,sompet2013dynamics,lester2015rapid,brown2019gray,jenkins2022ytterbium}. In both cases, however, the overall role of hyperfine structure is difficult to discern. A direct comparison between loss rates across multiple transitions, both red- and blue-detuned from resonance, is needed to distinguish the contribution of hyperfine interactions from competing effects in LACs. Such insights could enable precision control over laser-induced collisions in a variety of cold-atom experiments, and help identify atomic or molecular species for quantum-based technologies. 

In this Letter, we elucidate the influence of hyperfine-resolved molecular states in resonant dipole interactions by measuring LAC-induced two-body loss rates over the hyperfine spectrum of $^{87}\mathrm{Rb}$ for exactly two atoms co-trapped in an optical tweezer. To obtain our results, we introduce a collision-mediated diatomic imaging technique that exploits repulsive LACs to convert 2 atoms into 1 by controlling the kinetic energy gain of an atom pair possessing non-zero center-of-mass momentum. While this mechanism is commonly used for enhanced loading in optical tweezers, we show it can be used to circumvent two-atom collisional loss during fluorescence imaging \cite{schlosser2002collisional}, which typically requires alternative methods for detecting more than a single atom \cite{mcgovern2011counting,hilliard2015trap}. Finally, a model that incorporates molecular photoassociation potentials \cite{kemmann2004near} and the Landau-Zener formalism \cite{suominen1995optical} is used to understand the role of molecular state density and interaction strength in shaping the observed loss-rate spectrum.

\begin{figure}[b]
\includegraphics[width=\columnwidth]
{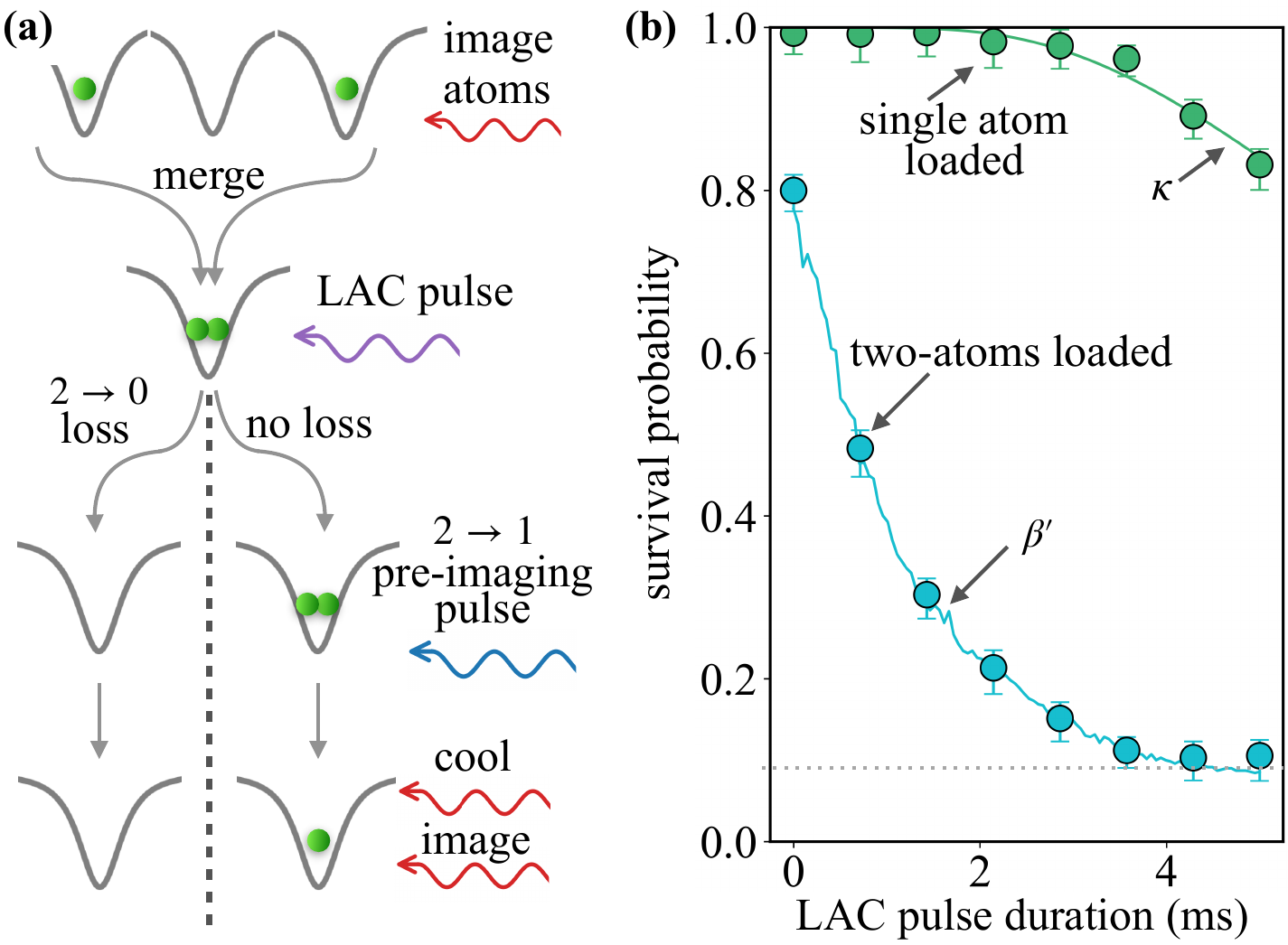}
\caption{\label{fig:experiment_diagram} (a) Experimental diagram for measuring two-body loss rates. An initial image is used to post-select instances when exactly two atoms are loaded. After merging atoms into a single tweezer, the LAC pulse is applied. The $2\!\rightarrow\! 1$ pre-imaging pulse removes one of the two remaining atoms if no loss occurred. (b) Example of post-selected two-atom (blue circles) and single-atom (green circles) survival probabilities with the corresponding two-body decay rate $\beta'$ and single atom heating rate $\kappa$. Here, collisional light is blue-detuned \SI{100}{MHz} from the D2 $F=2 \! \rightarrow\! F'=3$ transition.}
\end{figure}

Our experimental procedure [Fig.~\ref{fig:experiment_diagram}(a)] uses a $1 \! \times \!3$ array of optical tweezers generated with \SI{852}{\nm} light focused to a waist of $w_0$\SI{=700}{\nm}. Each tweezer is stochastically loaded with one $^{87}\mathrm{Rb}$ atom where an initial image allows us to post-select instances when exactly two atoms are loaded. The atoms are adiabatically merged into a single trap (see Appendix \ref{merge_appndx}) where the LAC pulse is applied for a variable duration. A $2 \!\! \rightarrow\!\! 1$ pre-imaging pulse is then used to remove a single atom if both atoms survive the initial collision, in which case the remaining atom is cooled and detected through fluorescence imaging on a CCD camera. The occurrence (absence) of a $2 \!\! \rightarrow\!\! 0$ loss event is therefore indicated by the presence of zero (one) atoms in the final image.

Atoms are loaded, imaged, and cooled to a temperature of $T=$ \SI{29(2)}{\uK} (measured via release and recapture \cite{tuchendler2008energy}) using $\sigma^+ \sigma^-$ polarization gradient cooling (PGC) red-detuned from the $F=2 \rightarrow F'=3$ transition, yielding the $|5^2 S_{1/2}, F \!=\! 2 \rangle$ state assumed to be in an approximately uniform $m_F$ distribution \cite{dalibard1989laser}. After merging into a single trap at a depth of $U/h\!=$ \SI{10(1)}{MHz}, collisional light is applied in conjunction with near-resonant D1 or D2 repump light at $I \approx 0.1 I_{\mathrm{sat}}$ to prevent population accumulation in $F\!=\!1$. The LAC pulse is collimated to a waist of \SI{1.4}{\mm} and applied as a $\pi$-polarized running wave along the radial dimension of the tweezers with an intensity of $I\! = \!1.2 I_{\mathrm{sat}}^{\mathrm{D1}}$ for D1 and $I\! = \!1.3 I_{\mathrm{sat}}^{\mathrm{D2}}$ for D2, where $I_{\mathrm{sat}}^{\mathrm{D1}} \! = \!$ \SI{4.5}{\frac{mW}{cm^2}} and $I_{\mathrm{sat}}^{\mathrm{D2}} \!=$ \SI{2.5}{\frac{mW}{cm^2}} \cite{steck2001rubidium}. This intensity is chosen to saturate the two-body loss rate (see Appendix \ref{intensity}), while the beam geometry (see Appendix \ref{beam_geo}) is chosen to avoid polarization gradient cooling or heating effects that can significantly alter atomic density prior to a collision event.

\begin{figure}[h!]
\includegraphics[width=\columnwidth]
{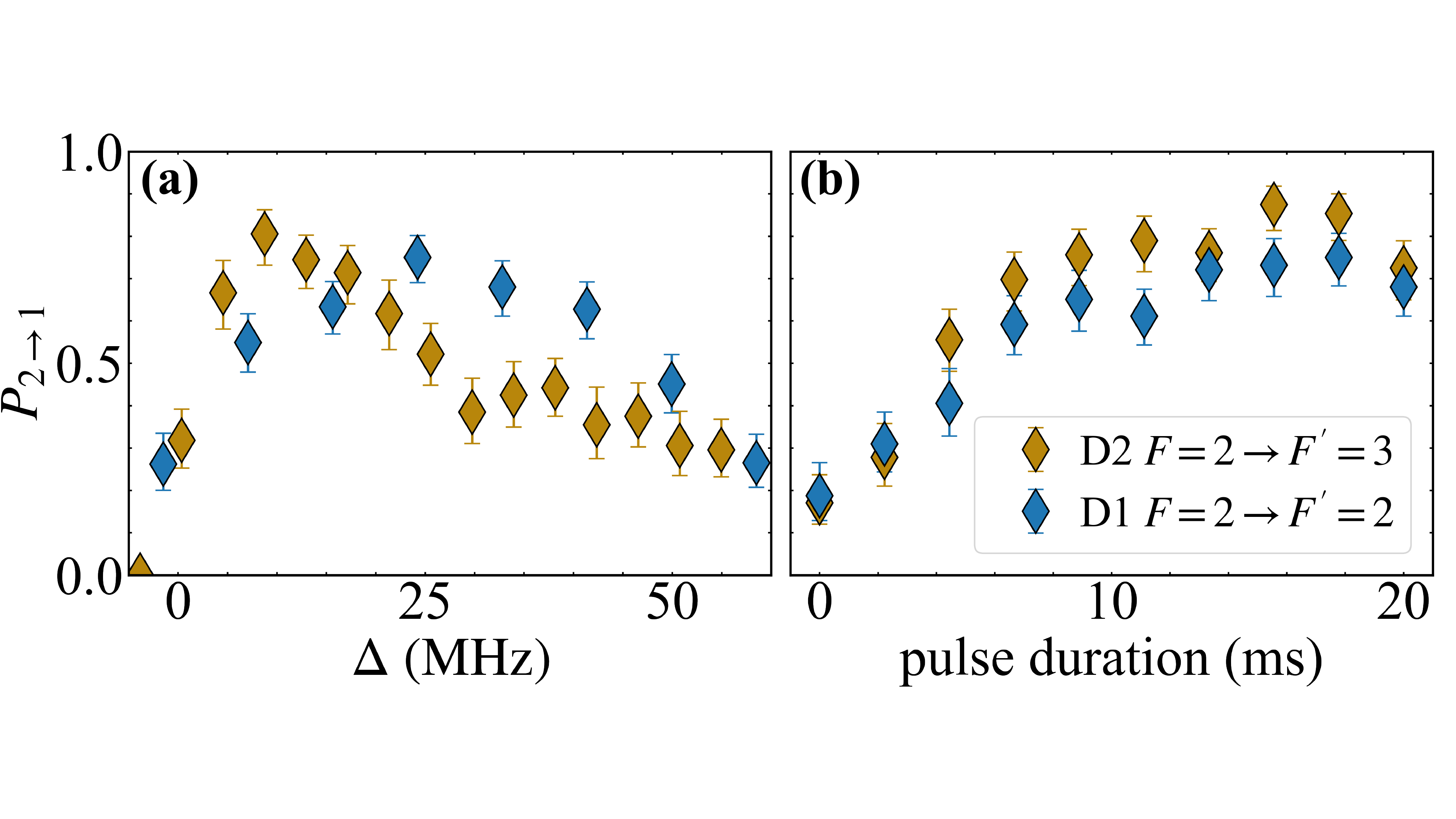}
\caption{\label{fig:2_to_1_probs} (a) $2 \! \rightarrow\! 1$ collision probability as a function of detuning  from the light-shifted ($U/h=$ \SI{10}{MHz}) resonance of the D1 $F=2 \! \rightarrow\! F'=2$  (blue) and D2 $F=2 \! \rightarrow\! F'=3 $ (gold) transitions using a \SI{15}{\ms} pulse duration with $I \approx I_{\mathrm{sat}}^{\mathrm{D1}}/5$ for D1 and  $I \approx I_{\mathrm{sat}}^{\mathrm{D2}}/50$ for D2. (b) $2 \! \rightarrow\! 1$ collision probability as a function of pulse duration for detunings of \SI{25}{MHz} (D1) and \SI{12}{MHz} (D2) ($I$ and $U$ same as above).   }
\end{figure}

The $2 \! \rightarrow\! 1$ pre-imaging pulse is focused through a \SI{100}{mm} achromatic lens to a waist of $w_0 \approx$ \SI{50}{\um} and applied along the radial direction of the tweezer while magnetic fields are zeroed at the atom location.  A maximum $80(6)$\% probability of converting two atoms into one atom ($P_{2 \! \rightarrow\! 1}$) is achieved at a trap depth of $U/h=$ \SI{10}{MHz} when using a \SI{15}{\ms} pulse and detuning $\Delta_{\mathrm{D2}} =$ \SI{12}{MHz} blue of the  light-shifted \footnote{All reported detunings $\Delta$ are accurate to $\pm$ \SI{4}{MHz} (see Appendix \ref{freespace_app}), and include a -12 MHz light shift from freespace resonance to account for the presence of the trap.} $F=2 \! \rightarrow\! F'=3$ transition. Although D2 light is exclusively used in the experimental procedure, a similar value of $P_{2 \! \rightarrow\! 1}$ can be achieved using light that is \SI{25}{MHz} blue-detuned of the D1  $F=2 \! \rightarrow\! F'=2 $ transition. This high conversion probability is not attainable on all hyperfine transitions, for instance, on the D1 $F=2 \! \rightarrow\! F'=1$ transition where we measure $P_{2 \! \rightarrow\! 1} \approx 20\%$. Figure \ref{fig:2_to_1_probs}(a) shows $P_{2 \! \rightarrow\! 1}$ versus detuning $\Delta$ from the D2 $F=2 \! \rightarrow\! F'=3$ (gold) and D1 $F=2 \! \rightarrow\! F'=2 $ (blue) transitions for a pulse duration of \SI{15}{ms}. Figure \ref{fig:2_to_1_probs}(b) shows $P_{2 \! \rightarrow\! 1}$ versus pulse duration at the D1 and D2 detunings mentioned above (note the non-zero value of $P_{2 \! \rightarrow\! 1}$ at zero pulse duration is due to $2 \! \rightarrow\!$ 1 collisions from the imaging light). Data in both plots is obtained when using a beam intensity of $ I \approx I_{\mathrm{sat}}^{\mathrm{D2}}/50$ for D2 and $I \approx I_{\mathrm{sat}}^{\mathrm{D1}}/5$ for D1. These values are empirically chosen to maximize $P_{2 \! \rightarrow\! 1}$, though the difference in intensity and detuning dependence between D1 and D2 is not well understood.

\begin{figure*}[t]
\includegraphics[width=\textwidth]{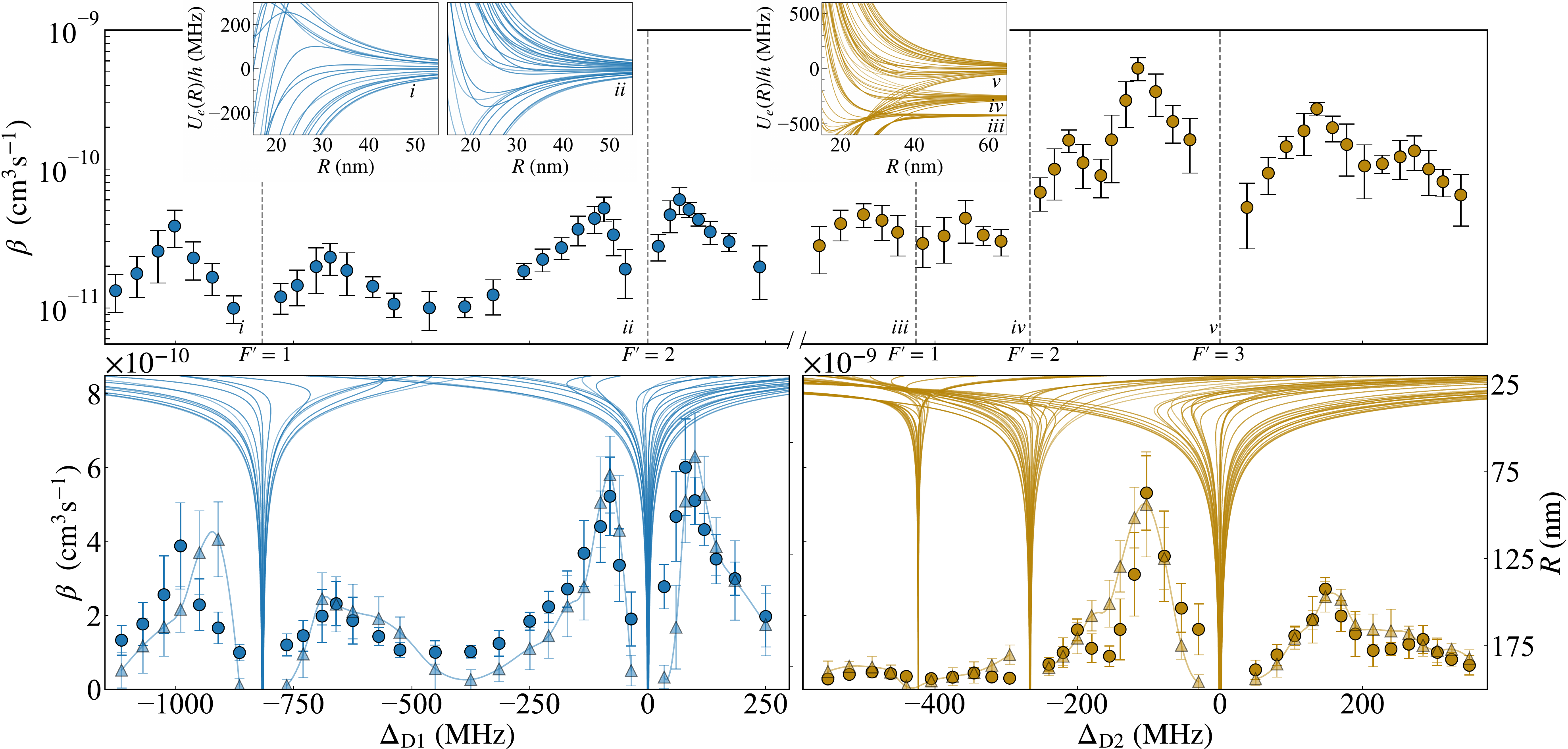}
\caption{\label{fig:beta_vs_detuning}
Two-body loss rates on a log (top) and linear (bottom) scale as a function of collisional light detuning ($\Delta_{\mathrm{D1}}, \Delta_{\mathrm{D2}}$) from the light-shifted ($U/h=$ \SI{10}{MHz}) resonances (dashed vertical lines) of the $F=2$ ground state to the hyperfine transitions (\textit{i-v}) of the D1 (blue circles) and D2 (gold circles) lines of $^{87}\mathrm{Rb}$. Intensity is fixed at $I\! = \!1.2  I_{\mathrm{sat}}^{\mathrm{D1}}$ for D1 and $I\! = \!1.3  I_{\mathrm{sat}}^{\mathrm{D2}}$ for D2. The insets (top) and right vertical axis (bottom) show calculated hyperfine-resolved molecular potentials $U_e(R)$ as a function of internuclear distance $R$, which are used to generate the simulated loss rates (triangles) from our model MC2. } 
\end{figure*}

The two-body loss rate is defined as 
$\beta = \beta'VN(N-1)$ where 
$V=\left({4 \pi k_b T}/{m\bar{\omega}^2}\right)^{3 / 2}$  
is the atomic volume with $\bar\omega=(\omega_r^2 \omega_a)^{1/3}$, and $N$ is the number of atoms \cite{kuppens2000loading}. We measure $\beta'$ by varying the collisional pulse duration and fitting the two-body decay profile with a Monte Carlo model (MC1) \cite{fuhrmanek2012light}. MC1 accounts for statistical contributions of the $2 \! \rightarrow\!1 $ pre-imaging pulse as well as single-atom heating rates that are measured via post-selection of data sets containing only one initial atom [Fig.~\ref{fig:experiment_diagram}(b)]. The $80(6)$\% upper limit of atom-pair survival probability is set by the $2 \! \rightarrow\! 1$ pre-imaging pulse, while the lower limit of 0.1-0.2 results from occasional $2\!\rightarrow\!1$ events driven by the LAC pulse or the imaging light. Both limits are accounted for in MC1, and found to be within the error bars in $\beta'$ (see Appendix \ref{tau_appndx}). Error bars in the atom-pair survival probability represent a $1-\sigma$ equal-tailed Jeffrey's prior confidence interval \cite{brown2001interval} over 200 repetitions, approximately 75 of which contain exactly two atoms. Error bars for $\beta$ include the corresponding uncertainty in $\beta^{\prime}$ and volume added in quadrature. 

Figure \ref{fig:beta_vs_detuning} shows two-body loss rates on a logarithmic (top) and linear (bottom) scale as a function of the LAC pulse detuning ($\Delta_{\mathrm{D1}},\Delta_{\mathrm{D2}}$) across the hyperfine transitions (\textit{i-v}) of the D1 (blue circles) and D2 (gold circles) lines of $^{87}\mathrm{Rb}$ (we omit the forbidden D2 $F=2 \rightarrow F'=0$ transition). The corresponding molecular photoassociation potentials $U_e(R)$ are calculated and plotted as a function of internuclear distance $R$, where each potential belongs to one of 14 molecular symmetries determined by the projection of total electronic and nuclear angular momentum along the internuclear axis \cite{jones2006ultracold}. To simplify our calculation, we ignore effects from rotation due to the small initial velocity of the atoms ($ \approx$ \SI{0.5}{cm/s}) and large interatomic distances at which collisions occur \cite{koch2012coherent}.

Triangles in the bottom plot show simulated two-body loss rates from a Monte Carlo model (MC2), with a global scaling factor applied. This model calculates the trajectories of two atoms in a Gaussian potential assuming a constant temperature of $T=$ \SI{29(2)}{\uK}. When atoms reach a Condon radius for a given detuning, the Landau-Zener transition probability $P_{LZ}=\exp \left (-2 \pi \hbar \Omega^2/\alpha|v_{rel}| \right )$  is used to calculate the inelastic collision probability \cite{suominen1995optical}. Here, $\Omega$ is the resonant electronic Rabi rate, $v_{rel}$ is the relative velocity of the atoms along the internuclear axis, and $\alpha\approx\left|\partial\left[U_e(R)\right]\partial R\right|_{R=R_C}$ is the slope parameter for any one of the 384 potentials. If a collision occurs, we solve the classical equations of motion for a single atom experiencing the potential for the duration of the excited-state lifetime. The kinetic energy gain is then evenly distributed between the two atoms along the internuclear axis. If the final kinetic energy of each atom exceeds the trap depth, a two-body loss event is recorded (additional details are provided in the Supplemental Material \cite{supplement}).

Modeling the hyperfine dependence of loss rates is complicated by unknown Rabi rates and excited-state lifetimes of potentials that conform to specific molecular symmetries, which may differ from those of the atomic dipole transition \cite{julienne1991cold}. To address this, we treat $\Omega$ as an adjustable parameter with a unique value for each hyperfine transition that is determined by reproducing the relative frequency and amplitude dependence of the measured loss rates. The values for $\Omega$, along with the corresponding single-atom values for oscillator strength, relative hyperfine intensity, and the branching ratios \cite{steck2001rubidium,ya2008quantum}, reproduce the absolute value of measured loss rates to within a global factor of 5 \cite{supplement}. 

The excited-state lifetime for each collision is randomly sampled from an exponential distribution corresponding to a decay rate of $\Gamma = 2\Gamma_a + \Gamma_{s}(\Delta)$, where $\Gamma_a$ is the natural linewidth of the atomic electronic dipole transition and $2 \Gamma_a$ is the molecular linewidth \cite{gallagher1989exoergic}. The stimulated emission rate $\Gamma_s(\Delta)$ suppresses excited-state lifetimes near resonance where most models becomes unreliable and overestimate loss on attractive potentials \cite{weiner2003cold}. This does not impact results for near-resonant blue detunings, where the kinetic energy gain is already insufficient for a single collision to yield loss. MC2 tends to underestimate loss rates in this regime by precluding the possibility of multiple collisions. Beyond the smallest detunings explored in this work, our model suggests that the loss rate $\beta$ and inelastic collision rate become equivalent as a single collision yields kinetic energy far beyond the trap depth.

The measured spectrum of Fig.~\ref{fig:beta_vs_detuning} exhibits loss-rate maxima $70$\SI{-200}{\MHz} red- and blue-detuned of each transition, with amplitudes spanning an order of magnitude. The detuning at which loss rates peak is believed to result from the balance between the inelastic collision probability and kinetic energy gain needed for atoms to escape the trap. In general, the kinetic energy gained in a LAC will be greater when the collision occurs at a smaller internuclear distance (or equivalently, larger detuning). However, atoms in a thermal state are less likely to reach small internuclear distances where highly energetic collisions can occur. This behavior is confirmed in MC2 and similar models of trap loss \cite{gallagher1989exoergic,julienne1991cold}. While the presence of bound states in attractive potentials can also yield an enhancement of loss rates \cite{miller1993photoassociation}, they are generally too dense to be experimentally resolvable for the detunings explored in this work \cite{weiner2003cold}. Our simulation also suggests that double-peak structures, like the one observed near $-200$ MHz of the D2 $F\!=\!2\!\rightarrow\!F'\!=\!3$ transition, can arise in regions where closely spaced transitions cause attractive and repulsive potentials to intersect.  

The maximum loss rate on the D2 line ($\sim$\SI{110}{MHz} red-detuned of  $F\!=\!2\!\rightarrow\!F'\!=\!3$) is about 8 times larger than the maximum D1 rate ($\sim$\SI{70}{MHz} blue-detuned of $F\!=\!2\! \rightarrow\! F'\!=\!2$). At least half of this difference can be attributed to single atom physics such as the dipole oscillator strength, which is twice as large for D2 than for D1. Branching ratios are also likely to play a role as the closed $F \!=\!2 \rightarrow\! F'\!=\!3$ transition effectively shields atoms from the $F\!=\!1$ ground state \cite{ya2008quantum} where two-body loss is less likely to occur (due to low intensity and near-resonant detuning of the repump light). In contrast, populations in $F'\!=\!2$ have an equal probability of decaying to the $F\!=\!2$  or $F\!=\!1$ ground state, yielding LACs at roughly half the rate of atoms in $F'\!=\!3$. Similarly, the relative strength of the D2 $F\!=\!2\!\rightarrow\!F'\!=\!3$ hyperfine transition is twice as large as the D1 $F\!=\!2\! \rightarrow\! F'\!=\!2$ \cite{steck2001rubidium}. 

The influence of hyperfine-dependent collisional effects is most unambiguously demonstrated by the two-fold difference in maximum rates between the D1 $F'=2$ and $F'=1$ transitions, where the only significant difference in single-atom physics is the branching ratios of $1/2$ and $5/6$ respectively \cite{ya2008quantum}. Scaling the loss-rate amplitude of each transition by the branching ratio, properties of the molecular potentials are estimated to contribute a factor of 10/3 to the relative difference in measured loss rates. MC2 indicates that a factor of 2 comes from the density of molecular states, where roughly 40 potentials exist for $F\!=\!2\! \rightarrow \! F'\!=\!2$ but only 10 exist for $F\!=\!2\! \rightarrow \! F'\!=\!1$ (thus providing fewer opportunities for collisions). The remaining contribution likely results from variations in molecular Rabi rates and lifetimes, as the average Condon radius is similar for both transitions.

 While the novelty of our results lie in the quantitative loss-rate variations across hyperfine transitions, the amplitude of absolute rates provides a useful indicator of single-atom loading timescales (see Appendix \ref{tau_appndx}) as well as for comparison to previous loss rate studies.  In the most comparable tweezer-based experiment \cite{fuhrmanek2012light}, loss rates measured near the D2  $F \!=\!2 \rightarrow\! F'\!=\!3$ transition were found to be 5 times larger than our reported value at a similar detuning of \SI{-30}{MHz}. This difference could be due to the high atomic temperatures or 3-body processes in Ref.~\cite{fuhrmanek2012light}. Other loss-rate experiments in tweezers \cite{schlosser2002collisional,sompet2013dynamics}, dipole traps \cite{gorges2008light}, and MOTs \cite{feng1993comparison,gensemer1997trap,jarvis2018blue} report loss-rate values anywhere from 2 to 5 times smaller than ours ($\beta$ from Refs.~\cite{schlosser2002collisional, sompet2013dynamics} are estimated). Direct comparisons between experiments are complicated by variations in spin-state distributions, as well as significant differences in detuning, intensity, and temperature. For example, when applying collisional light \SI{150}{MHz} blue-detuned of D1 $F\!=\!2\! \rightarrow \! F'\!=\!2$, we observe a factor of $2.5$ decrease in $\beta$ for atoms prepared at $T=$ \SI{50}{\uK} compared to $T=$ \SI{15}{\uK}. MC2 indicates that temperature dependence is a function of detuning, where an increase in temperature will lead to a decrease (increase) of $\beta$ at large (small) detunings. This is primarily due to a decrease in the relative distance between the average Condon radius and the radius corresponding to the atomic volume. However, the contribution of higher-order partial waves becomes important at hotter temperatures, resulting in a larger inelastic scattering cross-section that would further enhance loss rates.

In conclusion, our work sheds new light on the hyperfine dependence of resonant dipole interactions, and defines a quantitative benchmark for cold-atom/molecule applications where precision control over laser-induced collisions is advantageous.  This backdrop of information will be helpful in tracing hyperfine-resolved molecular-state contributions to enhanced sub-Poissonian loading, which for example, has been observed to work particularly well on the D1 $F\!=\!2 \!\rightarrow\! F'\!=\!2$ transition of $^{87}\mathrm{Rb}$ \cite{brown2019gray}. Furthermore, applications are not limited to only enhanced loading, but also number-resolved imaging ~\cite{mcgovern2011counting} and laser cooling processes in high density systems \cite{ruhrig2015high}. More broadly, harnessing methods demonstrated in this work will aid in comparing LACs among atomic species, 3-body systems, and even molecules.  

We would like to thank Paul Julienne, Antoine Browaeys, Mikkel Andersen, and Klaus M\"olmer for helpful discussions, as well as Zhenpu Zhang and Ting-Wei Hsu for their assistance. This work was supported by the Office of Naval Research (ONR) grant N00014-21-1-2594, National Science Foundation (NSF) PHY-2317149, NSF Quantum Leap Challenge Institutes (QLCI) award OMA - 2016244, the US Department of Energy, Office of Science, National Quantum Information Science Research Centers, Quantum Systems Accelerator, the Swiss National Science Foundation under grant 211072 (MM), and the Baur-SPIE Chair at JILA.

\begin{flushleft}
email: * steven.pampel@colorado.edu \\
email: $\dagger$ regal@colorado.edu
\end{flushleft}

\providecommand{\noopsort}[1]{}\providecommand{\singleletter}[1]{#1}%

\appendix

\section{Atom Merging Procedure}
\label{merge_appndx}
 Single atoms are loaded into 2 of the 3 available  tweezers with 35\% probability. After initial imaging and cooling, the tweezers are ramped to a trap depth of $U/h=$ \SI{1.9}{MHz} for the merging process. Here, the two outer tweezers are linearly ramped toward the central tweezer to achieve a final array spacing of \SI{440}{nm}, creating an effective single trap that now contains all atoms from the $1 \times 3$ array. The outer tweezers are then adiabatically removed to achieve the desired trap depth and frequency. Single- and two-atom temperatures measured before and after merging the traps indicate negligible heating is incurred in the process. The most-likely source of loss in the merging process comes from background gas collisions, which set our vacuum lifetime at \SI{4.8(1)}{s}. 

 \section{Intensity dependence}
\label{intensity}
\begin{figure}[h!]
\includegraphics[width=\columnwidth]
{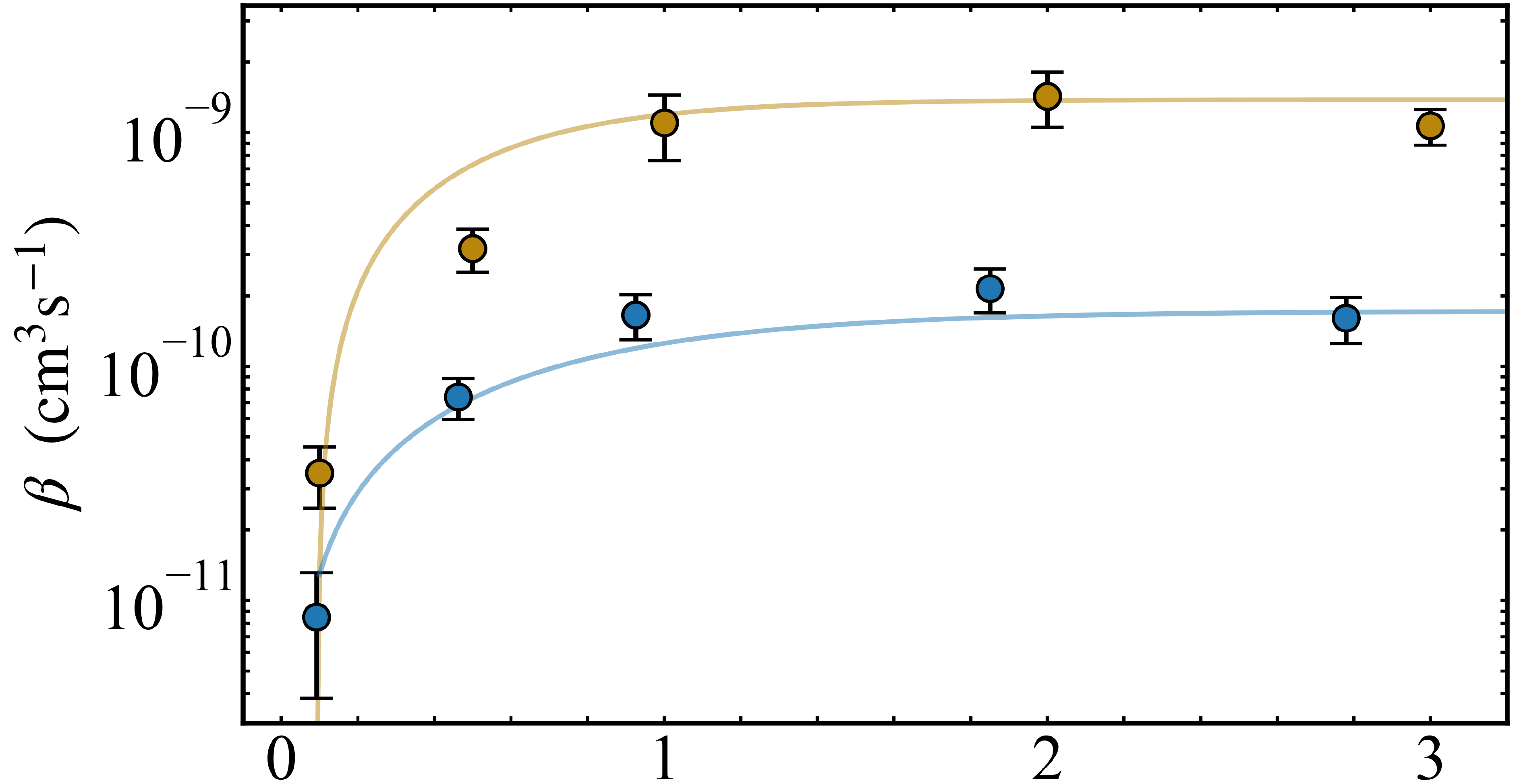}
\caption{\label{fig:intensity_fig} Two-body loss rates as a function of the LAC pulse saturation parameter $s$, where $\Delta_{\mathrm{D1}} = 65$ MHz blue-detuned of D1 $F\!=\!2 \!\rightarrow\! F'\!=\!2$ (blue circles), and $\Delta_{\mathrm{D2}} = 110$  MHz red-detuned of D2 $F\!=\!2 \!\rightarrow\! F'\!=\!3$ (gold circles). Corresponding fits are given by a tanh function.  }
\end{figure}

We examine loss rates as a function of intensity (Fig.~\ref{fig:intensity_fig}) with D1 light blue-detuned \SI{65}{MHz} of $F \!=\! 2 \rightarrow\! F'\! = \!2$ (blue circles) and D2 light red-detuned \SI{110}{MHz} of $F \!=\!2 \rightarrow\! F'\! = 3$ (gold circles). Light is applied in a polarization gradient cooling configuration to minimize effects of recoil and radiation pressure heating at different intensities.  The saturation parameter $s$ is defined as $I/I_{\mathrm{sat}}$ where $I_{\mathrm{sat}}^{\mathrm{D1}} \! = \!$ \SI{4.5}{\frac{mW}{cm^2}} for D1 and $I_{\mathrm{sat}}^{\mathrm{D2}} \! = \!$ \SI{1.6}{\frac{mW}{cm^2}}  for D2 \cite{steck2001rubidium}. Loss rates are observed to saturate close to $I_{\mathrm{sat}}$, which is consistent with previous studies \cite{fuhrmanek2012light,bali1994novel} and the intuition that inelastic scattering rates should be constrained by the dipole Rabi rate.

\section{Freespace resonance calibrations}
\label{freespace_app}
The D1 and D2 laser frequencies are stabilized via an offset lock from a reference laser locked to a Rb vapor cell. We calibrate the free-space resonance by measuring the in-trap resonance as a function of trap depth, where a linear fit of the data is used to extrapolate the frequency corresponding to freespace. Statistical uncertainty in the frequency calibration is added in quadrature with variation in the reference laser lock point, yielding a total uncertainty of $\pm$\SI{4}{MHz}. 

\section{Collisional Beam Geometry}
\label{beam_geo}
\begin{figure}[b]
\includegraphics[width=\columnwidth]
{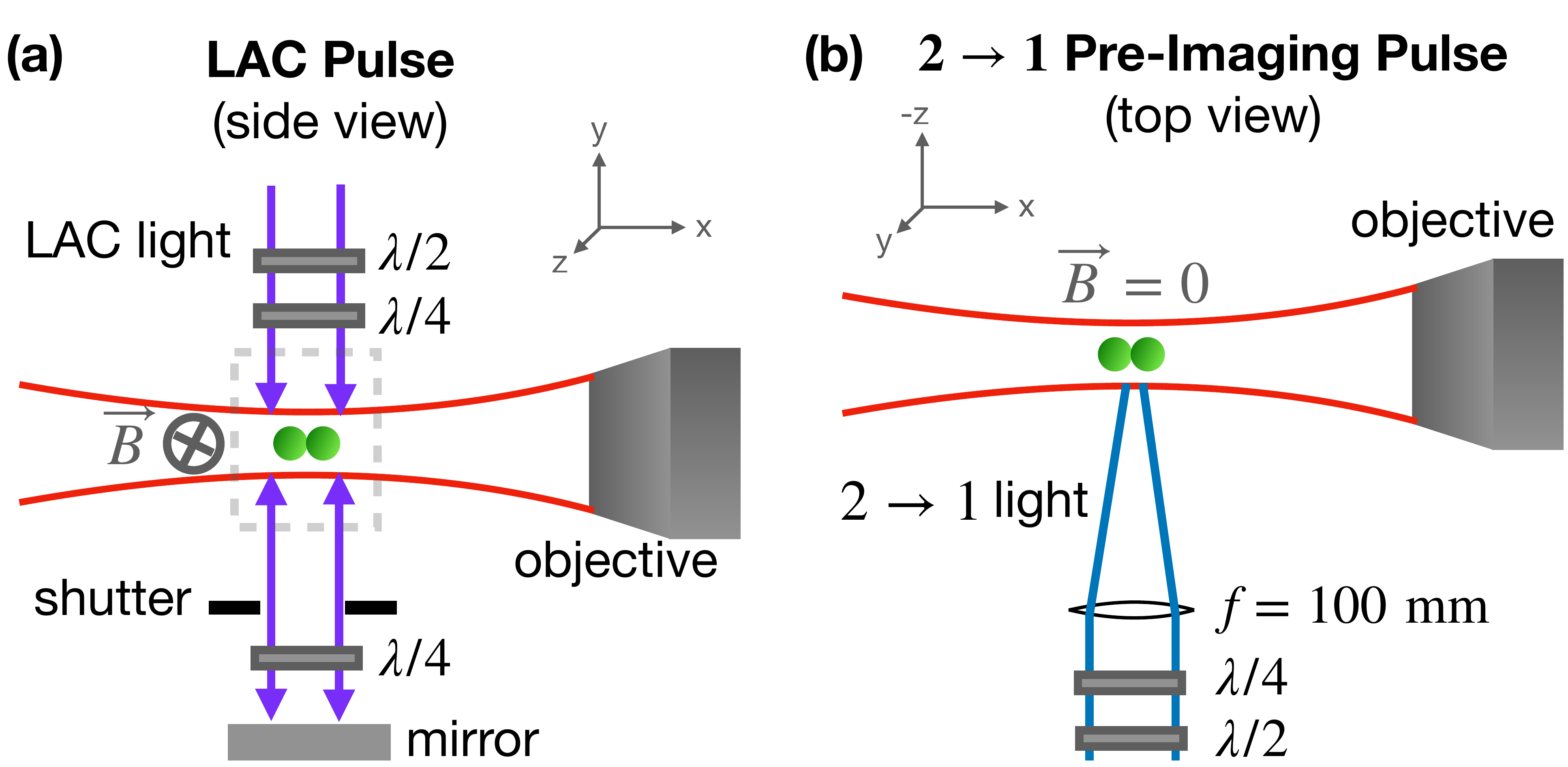}
\caption{\label{fig:beam_geometry} Beam geometry for the LAC pulse (a) and the $2 \! \rightarrow\! 1$ pre-imaging pulse (b). }
\end{figure}

The LAC pulse is collimated to a waist of \SI{1.4}{\mm} and applied in the $-y$ direction [Fig. \ref{fig:beam_geometry}(a)].  A shutter is used to block or transmit the back reflection of the LAC beam, enabling a running wave or standing wave condition for the LAC process. Only in the measurement of $\beta$ vs $s$ (Fig.~\ref{fig:intensity_fig}) do we keep the shutter open. D2 repump is applied along the diagonal MOT beam paths (not shown in the diagram), while D1 repump is applied along the LAC path. 

When a well-defined polarization is desired, an external magnetic field is applied in the $-z$ direction. To test for significant LAC polarization dependence, we compared loss rates using circularly ($\sigma^+ + \sigma^-$) and linearly ($\pi$) polarized collisional light red-detuned from the D2 $F=2 \rightarrow F'=3$ transition, and found no statistically meaningful difference. 

In the case of the $2 \rightarrow 1$ pre-imaging pulse, light is focused through a \SI{100}{mm} achromatic lens to a waist of $w_0 \approx$ \SI{50}{\um} and applied in a running wave configuration along the $-z$ direction [Fig. \ref{fig:beam_geometry}(b)]. Magnetic fields are zeroed at the atom location, however, this is not a necessary condition to drive $2 \rightarrow 1$ collisions with high probability. 

\section{1/e time vs detuning}
\label{tau_appndx}
\begin{figure*}[t]
\includegraphics[width=\textwidth]{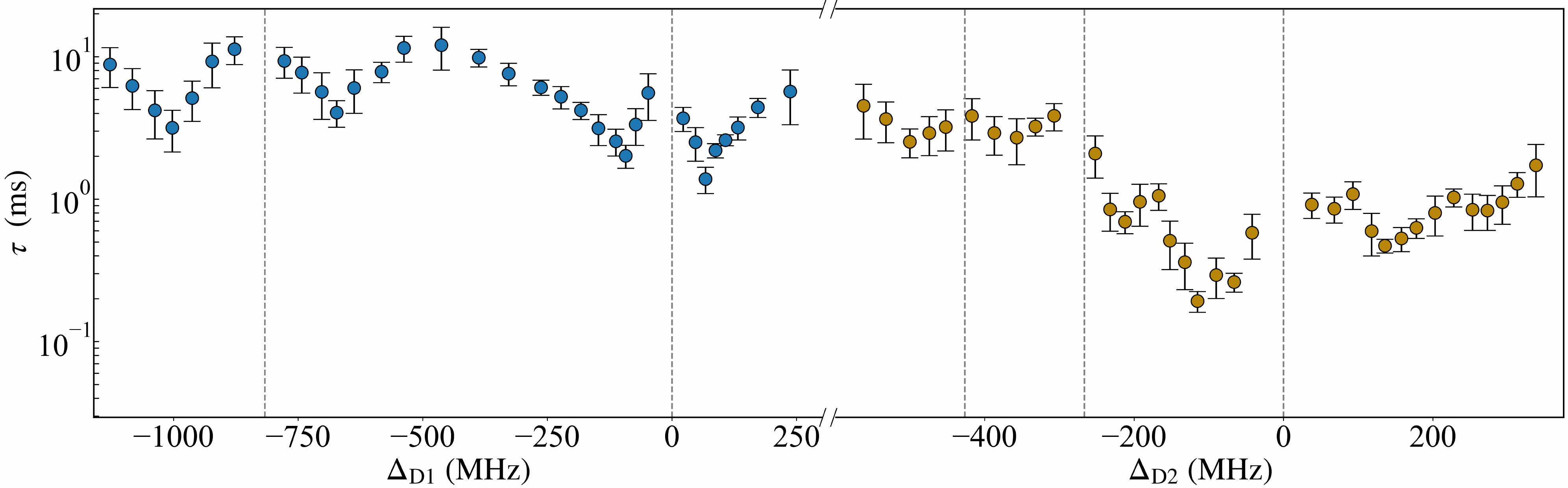}
\caption{\label{fig:tau_vs_detuning}
The 1/e time ($\tau$) of atom pair survival as a function of collisional light detuning across the hyperfine spectrum. Refer to caption of Fig.~\ref{fig:beta_vs_detuning} for trap and collisional light details. } 
\end{figure*}

 A complementary perspective to the loss-rate plot (Fig.~\ref{fig:beta_vs_detuning}) is shown in Fig.~\ref{fig:tau_vs_detuning}, where the 1/e time $\tau$ of atom pair survival is plotted as a function of the collisional light detuning across the hyperfine spectrum. Here,  $\tau= 1/\tilde{\beta}'$ where $\tilde{\beta}'$ is the raw two-body decay constant that is independent of MC1, and therefore does not account for single atom heating effects. Hence, $\tau$ near resonance is slightly smaller than Fig.~\ref{fig:beta_vs_detuning} would suggest. The very short timescale of atom-pair survival on the D2 $F \!=\! 2 \rightarrow\! F'\! = \!3$ transition is believed to be largely responsible for the fast single-atom loading times ($\sim 10-20$ ms). In contrast, we find loading on either of the D1 transitions, where the $\tau$ is much longer, generally takes on the order of $100$ ms. However, implications of absolute rates on enhanced loading timescales are a topic of ongoing research.  

\section*{Details of Numerical Model (MC2)}

We calculate the classical trajectory of two atoms in a Gaussian potential with initial position and velocity components randomly generated from a normal distribution spanning $\delta r_{x,y,z} = \pm \sqrt{k_B T/m \omega_{r,a}^2}$ and $\delta v_{x,y,z} = \pm \sqrt{k_B T/m}$, where $\omega_{r}$ ($\omega_{a}$) is the radial (axial) trap frequency. When atoms reach a resonant internuclear distance $R_C$ (Condon radius), where transitions to the $S+P$ molecular potentials occur, the Landau-Zener model is used to calculate the inelastic collision probability $P_R$ or $P_A$ for a repulsive or attractive potential $U_e(R)$ as \cite{weiner2003cold,sompet2013dynamics}:
\begin{equation}
P_{R}=2 P_{L Z}\left(1-P_{L Z}\right) \ \ \ \ \ \ 
P_{A}=1-\frac{P_{L Z}}{2-P_{L Z}}
\end{equation}
\begin{equation}
\label{P_LZ}
    P_{LZ}=\exp \left (-\frac{2 \pi \hbar \Omega^2}{\alpha|v_{rel}|} \right ).
\end{equation}  

\noindent Here, $\Omega = \gamma \Omega_J$ is the electronic Rabi rate where $\Omega_J=\sqrt{2d_J^2I_J/c \epsilon_0 \hbar^2}$ is the Rabi rate predicted by the D1 ($J=1/2$) or D2 ($J=3/2$) atomic dipole transition matrix element $d_J$ and LAC laser intensity $I_J$.  $\gamma$ is an adjustable parameter (see discussion of $\gamma$ below), $v_{rel}$ is the relative velocity of the atoms along the internuclear axis, and $\alpha\approx\left|\partial\left[U_e(R)\right]\partial R\right|_{R=R_C}$ is the slope parameter of the potential at $R_C$. We ignore interaction on the ground state potential, which is both standard in the Landau-Zener model and also a good approximation for long-range collisions. The different expressions for $P_R$ and $P_A$ result from the difference in available collisional pathways for attractive and repulsive potentials. In the case of repulsive potentials, atoms are either excited at $R_C$ when first approaching or moving away from one another. For attractive potentials, atoms can oscillate around $R_C$ an infinite number of times, resulting in a probability $P_A$ defined by the summation of each possible excitation \cite{weiner2003cold}.

Dipole-dipole interactions are typically studied with respect to a single potential defined by $U_e(R)=\pm C_3/R^3$, where $C_3$ is the dispersion coefficient that characterizes the strength of the electrostatic interaction. In this case, there is a single Condon radius for each detuning. In our case, where 384 potentials exist, each detuning can have dozens of Condon radii ranging from small to large internuclear distances. We determine the possible Condon radii by calculating the intersection of the detuning energy and molecular potentials of a given transition. The interaction region near a Condon radius defines where a collision can occur, and is determined by $\Delta R = |R_{+}-R_{-}|$, where $R_{\pm}$ is given by the the radial solutions of $U_e(R)= \hbar \Delta \pm \hbar \Omega/2$. Here, $\Delta$ is the collisional light detuning from a hyperfine resonance and $\Omega$ is the corresponding Rabi rate. The resulting interaction regions range from several nanometers (in the case of small detunings) to several picometers (in the case of large detunings or steep potentials). 

 When atoms reach an available Condon radius in the simulation, a collision occurs and we solve the classical equations of motion for a single atom experiencing the molecular potential. The final velocity, which depends on the initial conditions of the atoms, the potential being sampled, and the molecular lifetime, is evenly distributed between the two atoms along the internuclear axis. The final kinetic energy of each atom is then calculated to determine if it exceeds the depth of the trap. If a loss event is recorded, it is weighted by the inelastic collision probability to account for the contribution of Landau-Zener in the loss rate.  If a loss event does not occur, a new trajectory is initiated, therefore excluding the possibility of multiple collisions prior to a loss event. While multiple collisions are likely important at the smallest detunings, the slight improvement in accuracy was found to come at a large cost in simulation time. 

Our model makes several simplifying assumptions to reduce computational overhead. First, we assume constant temperature before a collision occurs, determined by randomly sampling from a normal distribution of the experimentally measured temperature and statistical uncertainty of \SI{29(2)}{\uK}.  We do not account for changes in momentum imparted from elastic scattering processes, and therefore assume a fixed oscillatory path defined by the initial conditions. In reality, the momentum vectors of the atoms are changing direction as they scatter photons and evolve in the trap. We approximate this effect by computing 80,000 trajectories at each detuning for a duration of the approximate oscillatory period of 5 $\mu$s. Due to the randomly generated initial positions, this approach also ensures that the atoms sample a wide variety of potentials occurring at different interatomic distances. By averaging the loss probability per time across all trajectories for a given detuning, we determine the two-body decay rate $\beta^{\prime}_{\mathrm{sim}}$ (defined in the same way as $\beta^{\prime}$ in Appendix E of the main text), which relates to the simulated two-body loss rate $\beta_{\mathrm{sim}}= \eta\beta^{\prime}_{\mathrm{sim}}  V N(N-1)$. Here, $V=1.1 \times 10^{-13} \mathrm{cm^3}$ is the atomic volume, $N=2$ is the number of atoms, and $\eta=5$ is a proportionality constant (see discussion of $\eta$ below). The error bars of $\beta_{\mathrm{sim}}$ represent a standard error over 8 simulations, each containing 10,000 trajectories.  Furthermore, we ignore the possibility of bound vibrational states and molecular-state coupling (fine/hyperfine-structure changing collisions and Stueckelberg oscillations), all of which have been shown to be negligible for the detunings explored in this work \cite{weiner2003cold, marcassa1999direct}. 

Accurately modeling collisional loss is complicated by unknown excited state lifetimes and Rabi rates for molecular potentials belonging to specific symmetry groups, which can vary significantly from those of the atomic dipole transition \cite{julienne1991cold}. To account for variations in the molecular Rabi rates, we treat $\gamma$ in the Rabi rate $\Omega=\gamma \Omega_J$ in Eq.~\ref{P_LZ} as an adjustable parameter to capture the relative frequency and amplitude dependence for each hyperfine transition. Optimization is accomplished by tuning the value of $\gamma$ (with a sampling resolution of 0.1) until the difference between the simulated and measured loss rates is minimized. The values of $\gamma$ (Table I) used to generate the simulation results (Fig.~4 of the main text) are constrained within a 68(5)\% confidence interval, and categorized by attractive ($\gamma_A$) and repulsive ($\gamma_R$) potentials. 

\begin{table}[h!]
    \centering
    \begin{tabular}{|c|c|c|c|c|c|}
        \hline
        \textbf{Transition} & \boldmath$\gamma_R$ & \boldmath$\gamma_A$ & \boldmath$I_{HF}$ & \textbf{BR}  & \boldmath$f_{osc}$\\
        \hline
        D1 $F\!=\!2\! \rightarrow \! F'\!=\!2$ &  $0.22(2)$ & $0.16(2)$  & $0.50$ & $0.84$ & $0.34$ \\
        \hline
        D1 $F\!=\!2\! \rightarrow \! F'\!=\!1$ & $0.20(2)$ & $0.12(1)$  & $0.50$ & $0.50$ & $0.34$ \\
        \hline
        D2 $F\!=\!2\! \rightarrow \! F'\!=\!3$ & $0.33(3)$ & $0.55(1)$ & $0.70$ & $1.00$ & $0.70$ \\
        \hline
        D2 $F\!=\!2\! \rightarrow \! F'\!=\!2$ & $0.20 (3)$ & $0.12(3)$ & $0.25$ & $0.50$ & $0.70$ \\
        \hline
        D2 $F\!=\!2\! \rightarrow \! F'\!=\!1$ & $0.12(3)$ & $0.15(3)$  & $0.05$ & $0.17$ & $0.70$ \\
        \hline
    \end{tabular}
    \caption{Values for the adjustable parameter $\gamma$ (constrained within a 68(5)\% confidence interval) in the molecular Rabi rate expression $\Omega= \gamma \Omega_J$ used in the simulated two-body loss rates (Fig.~4 of the main text) for attractive ($\gamma_A$) and repulsive ($\gamma_R$) potentials corresponding to each hyperfine transition on the D1 and D2 lines. Single-atom values for the relative hyperfine intensity $I_{HF}$ and dipole absorption oscillator strengths $f_{osc}$ are taken from Ref.~\cite{steck2001rubidium}, while values for the branching ratios (BR) are calculated from Ref.~\cite{ya2008quantum}.}
    \label{tab:rabi_table}
\end{table}

While $\gamma$ provides moderate control over the amplitude of the loss rates, it cannot be adjusted without also changing the frequency dependence. For the values shown in Table \ref{tab:rabi_table}, we find the absolute scale of $\beta_{\mathrm{sim}}$ is approximately a factor of 5 smaller than experiment, which is likely a result of simplifications within the model outlined in the previous paragraph. To account for this, we scale each value of $\beta_{\mathrm{sim}}$ by a global proportionality constant $\eta=5$. Consequently, MC2 only provides a rough estimate of the absolute Rabi rates; however, the relative differences between Rabi rates offer valuable insights into the influence of molecular potentials in shaping the loss-rate spectrum. 

To control $\Omega$ in the simulation, where potentials belonging to different transitions can overlap near a given detuning, we determine the dissociation limit belonging to the potential and apply corresponding Rabi rate for both the inelastic scattering probability $P_{R,A}$ and interaction region $\Delta R$. In addition to the molecular Rabi rate, properties relating to single atom-light interactions such as the relative hyperfine intensity \cite{steck2001rubidium}, absorption oscillator strength, and branching ratios (assuming a uniform distribution of $m_{F}$ states \cite{ya2008quantum}) are accounted for by scaling the inelastic collision probability for each transition by the appropriate factors from Table \ref{tab:rabi_table}.

To account for variations in excited-state lifetimes,  each lifetime is randomly sampled from a decaying exponential distribution corresponding to a linewidth of $\Gamma = 2\Gamma_a + \Gamma_{s}(\Delta)$. Here, $\Gamma_a$ is the natural linewidth of the atomic electronic dipole transition, $2\Gamma_a$ is natural molecular linewidth \cite{gallagher1989exoergic}, and the detuning-dependent stimulated emission rate is

\begin{equation}
 \Gamma_s(\Delta) = \frac{\pi \mu^2}{3 \epsilon_0 c \hbar^2} \frac{I(\Delta)}{1+ I(\Delta)/I_{\mathrm{sat}}} 
\end{equation}

\noindent with laser intenisty $I(\Delta)=\frac{\Gamma_a/2}{\Delta^2  +(\Gamma_a /2)^2} $. Effects from stimulated emission are expected to be important for near-resonant collisions \cite{gallagher1989exoergic}, providing strong suppression of the molecular-state lifetimes and corresponding two-body loss.  In this small-detuned regime, most models (including Landau-Zener) become unreliable and tend to over-estimate two-body loss on attractive potentials \cite{weiner2003cold}.

\end{document}